\documentclass[conference]{IEEEtran}
\IEEEoverridecommandlockouts
\usepackage{cite}
\usepackage{graphicx}
\usepackage[cmex10]{amsmath}
\usepackage{algorithmic}
\usepackage{algorithm}
\usepackage{array}
\usepackage{mdwmath}
\usepackage{mdwtab}
\usepackage[caption=false]{caption}
\usepackage{stfloats}

\ifCLASSOPTIONcaptionsoff
 \usepackage[nomarkers]{endfloat}
 \let\MYoriglatexcaption\caption
 \renewcommand{\caption}[2][\relax]{\MYoriglatexcaption[#2]{#2}}
\fi

\usepackage{url}
\begin{document}

\title{Doppler Spread Estimation by Subspace \\ Tracking for OFDM Systems}

\author{\IEEEauthorblockN{Xiaochuan~Zhao, Tao~Peng, Ming~Yang and Wenbo~Wang}%
\IEEEauthorblockA{Wireless Signal Processing and Network Lab \\
Key Laboratory of Universal Wireless Communication, Ministry of Education \\
Beijing University of Posts and Telecommunications, Beijing, China \\
Email: zhaoxiaochuan@gmail.com}%
\thanks{This work is sponsored in
part by the National Natural Science Foundation of China under grant
No.60572120 and 60602058, and in part by the national high
technology researching and developing program of China (National 863
Program) under grant No.2006AA01Z257 and by the National Basic
Research Program of China (National 973 Program) under grant
No.2007CB310602.}}

\maketitle

\begin{abstract}
In this paper, a novel maximum Doppler spread estimation algorithm
is presented for OFDM systems with the comb-type pilot pattern. The
least squared estimated channel frequency responses (CFR's) on pilot
tones are used to generate the auto-correlation matrices
with/without a known lag, from which the time correlation function
can be measured. The maximum Doppler spread is acquired by inverting
the time correlation function. Since the noise term will bias the
estimator, the estimated CFR's are projected onto the delay subspace
of the channel to reduce the bias term as well as the computation
complexity. Furthermore, the subspace tracking algorithm is adopted
to automatically update the delay subspace. Simulation results
demonstrate the proposed algorithm can quickly converge to the true
values for a wide range of SNR's and Doppler spreads in Rayleigh
fading channels.
\end{abstract}

\begin{IEEEkeywords}
Doppler spread, Estimation, Subspace tracking, OFDM, Time
correlation, Comb-type pilot.
\end{IEEEkeywords}

\IEEEpeerreviewmaketitle

\section{Introduction}
In order to cope with various radio transmission scenarios, adaptive
strategies, for example, adaptive modulation and coding and dynamic
resource allocation, are widely employed by many orthogonal
frequency division multiplexing (OFDM)-based wireless standards,
e.g., TAB, TVB, IEEE 802.11/16 and 3GPP LTE \cite{Ekst06}. Adaptive
schemes automatically adjust the system configurations and
transmission profiles according to some criteria to accommodate the
varying radio environments.

The maximum Doppler spread is one of the key parameters of criteria
for adaptive strategies. It determines the fading rate of the radio
channel and its reciprocal is a metric of the coherent time of the
channel. With the knowledge of it, wireless systems can change the
depths of interleavers to reduce coding/decoding latencies, decrease
unnecessary handoffs and adjust the rate of power control to reduce
signalling overhead. More important, for OFDM systems, when the
Doppler spread is comparable to the tone spacing, the orthogonality
between tones would be corrupted, which would arise the
inter-carrier interference (ICI) and consequently deteriorate the
system performance. However, if the Doppler spread can be estimated,
it will facilitate the adaptive schedule/control algorithms to
appropriately tune systems to alleviate ICI.

Most of existing methods of estimating the maximum Doppler spread
are categorized into two classes \cite{Tepe01}: level crossing rate
(LCR)-based and covariance (Cov)-based techniques. Since the
algorithms reviewed in \cite{Tepe01} were not specifically designed
for OFDM systems, they did not exploit the special signal structure
of OFDM systems. For OFDM systems, most literatures are Cov-based.
\cite{Scho02} determined the maximum Doppler spread through
estimating the smallest positive zero crossing point. In
\cite{Cai03}, Cai proposed to obtain the time auto-correlation
function (TACF) in time domain by exploiting the cyclic prefix (CP)
and its counterpart. However, Yucek \cite{Yuce05} pointed out that
for scalable OFDM systems whose CP sizes were varying over time,
\cite{Cai03} was difficult to offer a sufficient estimation of TACF,
which would degrade its estimation accuracy significantly. On the
contrary, Yucek proposed to estimate the Doppler spread through the
channel impulse responses (CIR's) which were estimated from the
periodically inserted training symbols.

Although the method in \cite{Yuce05} seems to be more robust than in
\cite{Cai03}, its shortcomings are also evident. For example, in
order to reduce system overheads, training symbols are arranged to
be far from each other, and typically transmitted as preambles to
facilitate the frame timing and carrier frequency synchronization.
Once the duration of frame is longer than the coherent time of the
channel, the maximum Doppler spread cannot be attained because TACF
turns to irreversible. Moreover, for sparse training symbols, the
converging speed of \cite{Yuce05} would be very slow, which hinders
its employment.

In this paper, we propose to estimate TACF by exploiting the
comb-type pilot tones \cite{Cole02} which are widely adopted in
wireless standards. In order to reduce noise perturbation, the
estimated channel frequency responses (CFR's) are projected onto the
delay subspace \cite{Sime04} to obtain CIR's, and the subspace
tracking algorithm \cite{Stro96} is adopted as well to track the
drifting delay subspace.

This paper is organized as follows. In Section \ref{sec:model}, the
OFDM system and channel model are introduced. Then, the maximum
Doppler spread estimation algorithm is presented in Section
\ref{sec:dopplerest}. Simulation results and analyses are provided
in Section \ref{sec:simulation}. Finally, Section
\ref{sec:conclusion} concludes the paper.

\setlength{\arraycolsep}{0.2em}

\section{System Model}
\label{sec:model} Consider an OFDM system with a bandwidth of
$BW=1/T$ Hz ($T$ is the sampling period). $N$ denotes the total
number of tones, and a CP of length $L_{cp}$ is inserted before each
symbol to eliminate inter-block interference. Thus the whole symbol
duration is $T_s = (N+L_{cp})T$. In each OFDM symbol, $P$ ($<N$)
tones are used as pilots to assist channel estimation. In addition,
optimal pilot pattern, i.e., equipowered and equispaced
\cite{Ohno04}, is assumed. Pilot indexes are collected in the set
${\mathcal{I}}_P$, i.e.,
${\mathcal{I}}_{P}=\{{\phi+p\times\theta}\}$, $(p=0,...,P-1)$, where
$\phi$ and $\theta$ are the offset and interval, respectively.

The discrete complex baseband representation of a multipath CIR of
length $L$ can be described by \cite{Stee92}
\begin{equation}
\label{eqn:discretecomplexmodel}
{h(n,\tau)=\sum\limits_{l=0}^{L-1}{\gamma_l(n)\delta\left({\tau-\tau_l}\right)}}\nonumber
\end{equation}
where $\tau_l$ is the delay of the $l$-th path, normalized by the
sampling period $T$, and $\gamma_l(n)$ is the corresponding complex
amplitude. Due to the motion of users, $\gamma_l(n)$'s are
wide-sense stationary (WSS) narrowband complex Gaussian processes,
and uncorrelated with each other based on the assumption of
uncorrelated scattering (US). In the sequel, $P\geq{L}$ and
$P\times\theta=N$ are assumed for determinability and simplicity,
respectively.

Furthermore, we assume the uniform scattering environment introduced
by Clarke \cite{Clar68}, thus $\gamma_l(n)$'s have the identical
normalized TACF $J_0(2\pi{f_d}t)$ for all $l$'s, where $f_d$ is the
maximum Doppler spread and $J_0(\cdot)$ is the zeroth order Bessel
function of the first kind. Hence, the discrete TACF is
\begin{equation}
\label{eqn:rtdef}
{r_{t,l}(m)=\sigma_l^2J_0\left(2\pi|m|{f_d}T\right)}
\end{equation}
where $\sigma _l^2$ is the power of the $l$-th path. Additionally we
assume the power of channel is normalized, i.e.,
$\sum\nolimits_{l=0}^{L-1}\sigma_l^2=1$.

Assuming a sufficient CP, i.e., $L_{cp} \geq L$, the signal model in
the time domain can be expressed as
\begin{equation}
\label{eqn:receivingsigdef}
{y_m(n)=\sum\limits_{l=0}^{L-1}h_m(n,\tau_l)x_m(n-\tau_l)+w_m(n)}\nonumber
\end{equation}
where $x_m(n)$ and $y_m(n)$ are the $n$-th samples of the $m$-th
transmitted and received OFDM symbols, respectively, $w_m(n)$ is the
sample of additive white Gaussian noise (AWGN), i.e.,
$E[w_m(n)w_m^*(n+q)]=\sigma_n^2\delta(q)$, and $h_m(n,\tau_l)$ is
the corresponding sample of the time-varying CIR. Through some
simple manipulations, the signal model in the frequency domain is
written as
\begin{equation}
\label{eqn:YmMatrixdef} {{\bf{Y}}_m={\bf{H}}_m{\bf{X}}_m+{\bf{W}}_m}
\end{equation}
where ${\bf{X}}_m,{\bf{Y}}_m, {\bf{W}}_m \in \mathcal{C}^{N\times1}$
are the $m$-th transmitted and received signal and noise vectors in
the frequency domain, respectively, and ${\bf{H}}_m \in
\mathcal{C}^{N\times{N}}$ is the channel transfer matrix whose
$(\nu+k,k)$-th element, i.e., $[{\bf{H}}_m]_{\nu+k,k}$, is
$\frac{1}{N}\sum\nolimits_{n=0}^{N-1}\sum\nolimits_{l=0}^{L-1}h_m(n,\tau_l)e^{-j2\pi(\nu{n}+k\tau_l)/N}$,
where $k$ denotes subcarrier while $\nu$ denotes Doppler spread.
Apparently, as ${\bf{H}}_m$ is non-diagonal, ICI is present.
However, when the normalized maximum Doppler spread, i.e., $f_dT_s$,
is less than 0.1, the signal-to-interference ratio (SIR) is over
17.8 dB \cite{Choi01}, which enables us to discard non-diagonal
elements of ${\bf{H}}_m$ with a negligible performance penalty.

As the comb-type pilot pattern is adopted, only pilot tones, denoted
as ${\bf{Y}}_{P;m}\in\mathcal{C}^{P\times1}$, are extracted from
${\bf{Y}}_m$. By approximating ${\bf{H}}_m$ to be diagonal,
(\ref{eqn:YmMatrixdef}) is modified to
\begin{equation}
\label{eqn:Ypmdef}
{{\bf{Y}}_{P;m}={\bf{X}}_{P;m}{\bf{H}}_{P;m}+{\bf{W}}_{P;m}}
\end{equation}
where ${\bf{X}}_{P;m}\in\mathcal{C}^{P\times{P}}$ is a pre-known
diagonal matrix, and ${\bf{H}}_{P,m}\in\mathcal{C}^{P\times1}$
consists of diagonal elements of ${\bf{H}}_m$. Hence, by denoting
the instantaneous CFR as
$H_m(n,k)=\sum\nolimits_{l=0}^{L-1}h_m(n,\tau_l)e^{-j2\pi{k\tau_l}/N}$,
we have
$[{\bf{H}}_{P;m}]_{p}=\frac{1}{N}\sum\nolimits_{n=0}^{N-1}H_m(n,\phi+p\times\theta)$.
Denote the Fourier transform matrix on the pilot tones as
${\bf{F}}_{P}\in\mathcal{C}^{P\times{N}}$, that is,
$[{\bf{F}}_P]_{p,n}=\frac{1}{\sqrt{N}}e^{-j2\pi(\phi+p\times\theta)n/N}$,
then ${\bf{W}}_{P;m}={\bf{F}}_{P}{\bf{w}}_m$, where
${\bf{w}}_m=[w_m(0),\ldots,w_m(N-1)]^T$, so,
$E[{\bf{W}}_{P;m}{\bf{W}}_{P;m}^H]=\sigma_n^2{\bf{I}}_P$.

\section{Maximum Doppler Spread Estimation}
\label{sec:dopplerest} At the receiver, the least-squared (LS)
channel estimation on pilot tones is carried out firstly, i.e.,
\begin{equation}
\label{eqn:Hpmestdef}
{{\bf{\hat{H}}}_{P;m}={\bf{X}}_{P;m}^{-1}{\bf{Y}}_{P;m}={\bf{H}}_{P;m}+{\bf{V}}_{P;m}}
\end{equation}
where ${\bf{\hat H}}_{P;m} \in \mathcal{C}^{P \times 1}$ is the
estimated CFR, and ${\bf{V}}_{P;m}\in \mathcal{C}^{P\times1}$ is the
noise vector expressed as
${{\bf{V}}_{P;m}={\bf{X}}_{P;m}^{-1}{\bf{W}}_{P;m}}$, hence,
${\bf{V}}_{P;m}\sim\mathcal{CN}(0,\sigma_n^2{\bf{I}}_P)$ when
${\bf{X}}_{P;m}^H{\bf{X}}_{P;m}={\bf{I}}_P$ for PSK modulated pilot
tones with equal power.

In the following,  we will introduce a method of estimating the
maximum Doppler spread based on TACF measured from significant paths
of the channel obtained by projecting the LS estimated CFR onto the
delay subspace.

\subsection{Measurement of the Time Auto-Correlation Function}
First, by defining the Fourier transform matrix as
${\bf{F}}_{P,\tau}\in\mathcal{C}^{P\times{L}}$ with
$[{\bf{F}}_{P,\tau}]_{p,l}=e^{-j2\pi(\phi+p\times\theta)\tau_l/N}$,
${\bf{H}}_{P;m}$ can be expressed as
\begin{equation}
{\bf{H}}_{P;m}=\frac{1}{N}\sum\limits_{n=0}^{N-1}{\bf{H}}_{P;m}(n)=\frac{1}{N}\sum\limits_{n=0}^{N-1}{\bf{F}}_{P,\tau}{\bf{h}}_{m}(n)\nonumber
\end{equation}
where ${\bf{H}}_{P;m}$ and ${\bf{h}}_{m}(n)$ are CFR and
instantaneous CIR vectors, respectively. Regardless of noise, the
0-lag auto-correlation matrix of ${\bf{H}}_{P;m}$ is
\begin{eqnarray}
{\bf{R}}_{{{\bf{H}}_P}}(0)&=&E\left[{\bf{H}}_{P;m}{\bf{H}}_{P;m}^H\right]\nonumber\\
\label{eqn:avgHdef}
{}&=&\frac{1}{N^2}\sum\limits_{n=0}^{N-1}\sum\limits_{q=0}^{N-1}{\bf{F}}_{P,\tau}{\bf{A}}_m(n,q){\bf{F}}_{P,\tau}^H
\end{eqnarray}
where ${\bf{A}}_m(n,q)=E[{\bf{h}}_{m}(n){\bf{h}}_{m}^H(q)]$, and
based on the assumption of WSSUS and Clarke model, its
$(l_1,l_2)$-th element is
$\left[{\bf{A}}_m(n,q)\right]_{l_1,l_2}=\sigma_{l_1}^2r_t(n-q)\delta(l_1-l_2)$,
where $r_t(n)$ is the normalized TACF, hence ${\bf{A}}_m(n,q)$ is
diagonal. Denoting ${\bf{D}}=diag(\sigma_l^2)$, $l=0,\ldots,L-1$, we
have
\begin{equation}
\label{eqn:Adef} {\bf{A}}_m(n,q)=r_t(n-q){\bf{D}}
\end{equation}
Substitute (\ref{eqn:Adef}) into (\ref{eqn:avgHdef}) and with some
manipulations
\begin{eqnarray}
\label{eqnarray:Rhp0def}
{\bf{R}}_{{{\bf{H}}_P}}(0)&=&\xi(0){\bf{F}}_{P,\tau}{\bf{D}}{\bf{F}}_{P,\tau}^H\\
\label{eqnarray:xi0def}
\xi(0)&=&\frac{1}{N^2}\sum\limits_{n=0}^{N-1}\sum\limits_{q=0}^{N-1}r_t(n-q)
\end{eqnarray}
Similarly, the $\beta$-lag auto-correlation matrix of
${\bf{H}}_{P;m}$ ($\beta\ge0$), defined as
${\bf{R}}_{{{\bf{H}}_P}}(\beta)=E[{\bf{H}}_{P;m+\beta}{\bf{H}}_{P;m}^H]$,
can be written as
\begin{eqnarray}
\label{eqnarray:Rhpbetadef}
{\bf{R}}_{{{\bf{H}}_P}}(\beta)&=&\xi(\beta){\bf{F}}_{P,\tau}{\bf{D}}{\bf{F}}_{P,\tau}^H \\
\label{eqnarray:xibetadef}
\xi(\beta)&=&\frac{1}{N^2}\sum\limits_{n=0}^{N-1}\sum\limits_{q=0}^{N-1}r_t(n-q+(N+L_{cp})\beta)
\end{eqnarray}
Then, with (\ref{eqnarray:Rhp0def}) and (\ref{eqnarray:Rhpbetadef}),
we have
\begin{equation}
\label{eqn:etadef}
\eta=\sqrt{\frac{||{\bf{R}}_{{\bf{H}}_P}(\beta)||_F^2}
{||{\bf{R}}_{{\bf{H}}_P}(0)||_F^2}}=\frac{\xi(\beta)}{\xi(0)}
\end{equation}
where $||{\cdot}||_F$ denotes the Frobenius norm. When the
normalized Doppler spread $f_dT_s\le0.1$, referring to
(\ref{eqn:rtdef}), we can make an approximation (which we will
examine later) as
\begin{equation}
\label{eqn:etaapprox} \eta\;\approx\;J_0(2\pi\beta(N+L_{cp})f_dT)
\end{equation}
Since when $\beta(N+L_{cp})f_dT=\beta f_dT_s\leq0.38$, $J_0(\cdot)$
is positive and reversible, meanwhile, in order to hold the
orthogonality between subcarriers, $f_dT_s$ is usually smaller than
0.1, thus $\beta\le3$ is the feasible range. Then $f_d$ can be
estimated by
\begin{equation}
\label{eqn:estfd} \hat{f_d} = \frac{J_0^{-1}(\eta)}{2\pi\beta{T_s}}
\end{equation}

Now we consider the effect of noise. When noise is present,
(\ref{eqnarray:Rhp0def}) and (\ref{eqnarray:Rhpbetadef}) are
rewritten into
\begin{eqnarray}
\label{eqnarray:Rhp0defnew}
{\bf{\hat{R}}}_{{{\bf{H}}_P}}(0)&=&\xi(0){\bf{F}}_{P,\tau}{\bf{D}}{\bf{F}}_{P,\tau}^H+\sigma_n^2{\bf{I}}_P\\
\label{eqnarray:Rhpbetadefnew}
{\bf{\hat{R}}}_{{{\bf{H}}_P}}(\beta)&=&\xi(\beta){\bf{F}}_{P,\tau}{\bf{D}}{\bf{F}}_{P,\tau}^H
\end{eqnarray}
Correspondingly, (\ref{eqn:etadef}) changes into
\begin{equation}
\label{eqn:etadefnew}
\eta=\sqrt{\frac{||{\bf{\hat{R}}}_{{\bf{H}}_P}(\beta)||_F^2}
{||{\bf{\hat{R}}}_{{\bf{H}}_P}(0)||_F^2}}=\sqrt{\frac{\xi(\beta)^2}{\xi(0)^2+\rho^2}}
\end{equation}
where $\rho$ is defined as
\begin{equation}
\label{eqn:rhodef}
\rho=\sqrt{\frac{P\sigma_n^4}{||{\bf{F}}_{P,\tau}{\bf{D}}{\bf{F}}_{P,\tau}^H||_F^2}}
\end{equation}
As pilot tones are equispaced,
${\bf{F}}_{P,\tau}^H{\bf{F}}_{P,\tau}=P{\bf{I}}_L$, then
\begin{equation}
||{\bf{F}}_{P,\tau}{\bf{D}}{\bf{F}}_{P,\tau}^H||_F^2=P^2\sum\limits_{l=0}^{L-1}\sigma_l^4\nonumber
\end{equation}
therefore, (\ref{eqn:rhodef}) is
\begin{equation}
\label{eqn:rhodefnew}
\rho=\sqrt{\frac{\sigma_n^4}{P\sum\nolimits_{l=0}^{L-1}\sigma_l^4}}
\end{equation}
Note
$\sum\nolimits_{l=0}^{L-1}\sigma_l^4\le(\sum\nolimits_{l=0}^{L-1}\sigma_l^2)^2$,
we have
\begin{equation}
\label{eqn:rholowbound} \rho\ge\frac{1}{\sqrt{P}\times{SNR}}
\end{equation}
where $SNR=\sigma_n^{-2}$ for the normalized power of the channel
and pilot tones.

\subsection{Improving the Estimation Accuracy by the Delay Space}
Although the maximum Doppler spread can be evaluated from
(\ref{eqn:etadefnew}) and (\ref{eqn:estfd}), the effect of noise
will bias the result of estimation heavily when $P$ is small and SNR
is low. On the other hand, when $P$ is large, the effect of noise is
negligible, but the sizes of ${\bf{\hat{R}}}_{{\bf{H}}_P}(0)$ and
${\bf{\hat{R}}}_{{\bf{H}}_P}(\beta)$ turn to be so large that
evaluating their Frobenius norms would require a large amount of
calculations, which hinders the employment of this method in real
applications. Therefore, we introduce the delay subspace onto which
the auto-correlation matrices are projected to reduce the effect of
noise as well as the computation complexity.

Firstly, the eigenvalue decomposition (EVD) is performed
\begin{eqnarray}
\label{eqnarray:R0evd}
{\bf{\hat{R}}}_{{{\bf{H}}_P}}(0)&=&{\bf{U}}_{\tau}\left[\xi(0){\bf{\Lambda}}+\sigma_n^2{\bf{I}}_P\right]{\bf{U}}_{\tau}^H\\
\label{eqnarray:Rbetaevd}
{\bf{\hat{R}}}_{{{\bf{H}}_P}}(\beta)&=&{\bf{U}}_{\tau}\left[\xi(\beta){\bf{\Lambda}}\right]{\bf{U}}_{\tau}^H
\end{eqnarray}
Since the number of channel taps is $L$,
$rank({\bf{\Lambda}})=L\le{P}$, the last $P-L$ eigenvalues of
${\bf{\hat{R}}}_{{{\bf{H}}_P}}(0)$ and
${\bf{\hat{R}}}_{{{\bf{H}}_P}}(\beta)$ are $\sigma_n^2$ and 0,
respectively. Once $L$ is available, ${\bf{U}}_{\tau}$ can be
divided into two parts named as the "signal" and "noise" subspaces,
respectively, i.e.,
${\bf{U}}_{\tau}=[{\bf{U}}_{\tau,s},{\bf{U}}_{\tau,n}]$, where
${\bf{U}}_{\tau,s}\in\mathcal{C}^{P\times{L}}$, and so does
${\bf{\Lambda}}$, i.e.,
${\bf{\Lambda}}=diag({\bf{\Lambda}}_{s},{\bf{0}}_{P-L})$, where
${\bf{\Lambda}}_{s}\in\mathcal{C}^{L\times{L}}$. Hence,
\begin{eqnarray}
\label{eqnarray:subspaceR0}
{\bf{U}}_{\tau,s}^H{\bf{\hat{R}}}_{{{\bf{H}}_P}}(0){\bf{U}}_{\tau,s}&=&\xi(0){\bf{\Lambda}}_{s}+\sigma_n^2{\bf{I}}_L\\
\label{eqnarray:subspaceRbeta}
{\bf{U}}_{\tau,s}^H{\bf{\hat{R}}}_{{{\bf{H}}_P}}(\beta){\bf{U}}_{\tau,s}&=&\xi(\beta){\bf{\Lambda}}_{s}
\end{eqnarray}

Based on (\ref{eqnarray:subspaceR0}) and
(\ref{eqnarray:subspaceRbeta}), (\ref{eqn:etadefnew}) can be refined
as
\begin{equation}
\label{eqn:etadefnewrefine}
\eta=\sqrt{\frac{||{\bf{U}}_{\tau,s}^H{\bf{\hat{R}}}_{{{\bf{H}}_P}}(\beta){\bf{U}}_{\tau,s}||_F^2}
{||{\bf{U}}_{\tau,s}^H{\bf{\hat{R}}}_{{{\bf{H}}_P}}(0){\bf{U}}_{\tau,s}||_F^2}}=\sqrt{\frac{\xi(\beta)^2}{\xi(0)^2+\rho_r^2}}
\end{equation}
where $\rho_r$ is defined as
\begin{equation}
\label{eqn:rhodefnewrefine}
\rho_r=\sqrt{\frac{L\sigma_n^4}{||{\bf{\Lambda}}_{s}||_F^2}}
\end{equation}
From
(\ref{eqnarray:Rhp0defnew})(\ref{eqnarray:Rhpbetadefnew})(\ref{eqnarray:R0evd})(\ref{eqnarray:Rbetaevd}),
\begin{equation}
||{\bf{F}}_{P,\tau}{\bf{D}}{\bf{F}}_{P,\tau}^H||_F^2=||{\bf{\Lambda}}_{s}||_F^2\nonumber
\end{equation}
Hence, comparing (\ref{eqn:rhodefnewrefine}) with
(\ref{eqn:rhodef}), the bias term is reduced because
\begin{equation}
\label{eqn:rhorandrho} \frac{\rho_r}{\rho}=\sqrt{\frac{L}{P}}\le1
\end{equation}

Actually, in the real circumstance, the number of significant taps
of wireless channels is far less than of pilot tones, thereby
projecting auto-correlation matrices onto the delay subspace, like
(\ref{eqnarray:subspaceR0}) and (\ref{eqnarray:subspaceRbeta}), can
effectively reduce the bias term and ease the calculation of $\eta$.

\subsection{Tracking the Delay Subspace -- the Proposed Algorithm}
When the user is moving, the tap delays of the channel, i.e.,
$\tau_l$'s, are slowly drifting \cite{Tse05}\cite{Sime04}, which
causes ${\bf{F}}_{P;\tau}$ to vary and so does ${\bf{U}}_{\tau,s}$.
To accommodate this variation, the subspace tracking algorithm is
adopted to automatically update the delay subspace. In addition, if
the number of significant taps of the channel is unknown, minimum
description length (MDL) \cite{Wax85} is employed to estimate it.
The proposed algorithm is summarized as follows.
\begin{algorithm}
\begin{algorithmic}

\STATE{\tt\small \textbf{Initialize}: ($n=0$) \vspace{3pt}\\
\begin{array}{l@{\;=\;}l@{\;=\;}l}
{\bf{Q}}_{0}(0)&{\bf{Q}}_{\beta}(0)&[{\bf{I}}_{L_m},{\bf{0}}_{L_m,P-L_m}^T]^T\nonumber\\
{\bf{A}}_{0}(0)&{\bf{A}}_{\beta}(0)&{\bf{0}}_{P,L_m}\nonumber\\
{\bf{C}}_{0}(0)&{\bf{C}}_{\beta}(0)&{\bf{I}}_{L_m}\nonumber
\end{array} \vspace{3pt}\\
\vspace{3pt}}

\STATE{\tt\small \textbf{Run}: ($n=n+1$)\vspace{3pt}\\
\parindent 2mm Input: ${\bf{\hat{H}}}_{P}(n)$\vspace{3pt}\\
\parindent 2mm 1) Updating for the 0-lag auto-correlation matrix:\vspace{3pt}\\
\begin{array}{l@{\;=\;}l}
{\bf{Z}}_{0}(n)&{\bf{Q}}_{0}(n-1){\bf{\hat{H}}}_{P}(n)\nonumber\\
{\bf{A}}_{0}(n)&\alpha{\bf{A}}_{0}(n-1){\bf{C}}_{0}(n-1)+(1-\alpha){\bf{\hat{H}}}_{P}(n){\bf{Z}}_{0}(n)^H\nonumber\\
{\bf{A}}_{0}(n)&{\bf{Q}}_{0}(n){\bf{R}}_{0}(n) : \textit{\textbf{QR-factorization}}\nonumber\\
{\bf{C}}_{0}(n)&{\bf{Q}}_{0}(n-1)^H{\bf{Q}}_{0}(n)\nonumber\\
{\hat{L}}(n)&MDL\left(diag({\bf{R}}_{0}(n))\right)\nonumber
\end{array} \vspace{3pt}\\

\parindent 2mm 2) Updating for the $\beta$-lag auto-correlation matrix:\vspace{3pt}\\
\begin{array}{l@{\;=\;}l}
{\bf{Z}}_{\beta}(n)&{\bf{Q}}_{\beta}(n-1){\bf{\hat{H}}}_{P}(n-\beta)\nonumber\\
{\bf{A}}_{\beta}(n)&\alpha{\bf{A}}_{\beta}(n-1){\bf{C}}_{\beta}(n-1)+(1-\alpha){\bf{\hat{H}}}_{P}(n){\bf{Z}}_{\beta}(n)^H\nonumber\\
{\bf{A}}_{\beta}(n)&{\bf{Q}}_{\beta}(n){\bf{R}}_{\beta}(n) : \textit{\textbf{QR-factorization}}\nonumber\\
{\bf{C}}_{\beta}(n)&{\bf{Q}}_{\beta}(n-1)^H{\bf{Q}}_{\beta}(n)\nonumber
\end{array} \vspace{3pt}\\

\parindent 2mm 3) Estimating $\eta$ according to (\ref{eqn:etadefnewrefine}):\vspace{3pt}\\
\begin{equation}
\hat{\eta}=\sqrt{\frac{\sum\nolimits_{l=1}^{{\hat{L}}(n)}|\left[{\bf{R}}_{\beta}(n)\right]_{l,l}|^2}{\sum\nolimits_{l=1}^{{\hat{L}}(n)}|\left[{\bf{R}}_{0}(n)\right]_{l,l}|^2}}\nonumber
\end{equation} \vspace{3pt}\\

\parindent 2mm 4) Estimating $f_d$ according to (\ref{eqn:estfd}).\vspace{3pt}\\
}

\STATE{\tt\small \textbf{Remark}: $\alpha$ is a positive exponential
forgetting factor close to 1. $L_{m}$ is the maximum rank to be
tested. $MDL(\cdot)$ denotes the MDL detector and $diag({\bf{R}})$
denotes the diagonal elements of ${\bf{R}}$. In the simulation, we
set $\alpha=0.995$ and $L_m=10$.}
\end{algorithmic}
\end{algorithm}

\subsection{Other Considerations}
In this subsection, several further discussions about the proposed
algorithm are presented.

First, numerical results are shown in Table \ref{tab:etaapprox} to
examine (\ref{eqn:etaapprox}) when $N=512$, $L_{cp}=64$ and
$\beta=1$. From Table \ref{tab:etaapprox}, we find
(\ref{eqn:etaapprox}) is a good approximation when $\beta{f_dT_s}$
is small. It is worth noting that
${J_0(2\pi\beta{f_d}T_s)}\le{\eta}$, hence
$f_d\ge\frac{J_0^{-1}(\eta)}{2\pi\beta{T_s}}$, in other words,
(\ref{eqn:etaapprox}) tends to over-estimate the maximum Doppler
spread a bit.

\begin{table}[t]
\renewcommand{\arraystretch}{2}
\caption{A Table of Values of (\ref{eqn:etaapprox})}
\label{tab:etaapprox} \centering
\begin{tabular}{|c|c|c|c|c|}
\hline
$f_dT_s$ & 0.02 & 0.04 & 0.06 & 0.08 \\
\hline
$J_0(2\pi\beta{f_d}T_s)$ & 0.9961 & 0.9843 & 0.9648 & 0.9378 \\
\hline
$\eta\left(=\frac{\xi(\beta)}{\xi(0)}\right)$ & 0.9961 & 0.9843 & 0.9649 & 0.9381 \\
\hline
$\frac{J_0(2\pi\beta{f_d}T_s)}{\eta}$ & 1.0000 & 1.0000 & 0.9999 & 0.9997 \\
\hline
\end{tabular}
\end{table}

Then we compare the computation complexity of the proposed algorithm
with (\ref{eqn:etadefnew}). The computation complexity of the
subspace tracking is $\mathcal{O}(P\times{L}^2)$ \cite{Stro96}.
Since it takes a dominant proportion of the total number of
instructions required by the proposed algorithm, we use it instead.
Meanwhile, the computation complexity of (\ref{eqn:etadefnew}),
which directly computes the Frobenius norm of $P\times{P}$ matrices,
is $\mathcal{O}(P^2)$. Apparently, when $P>{L}^2$, which is usually
the case for sparse multipath channels, the proposed algorithm can
reduce the computation complexity considerably.

\begin{figure}[!t]
\centering
\includegraphics[width=3.7in, height=2.3in]{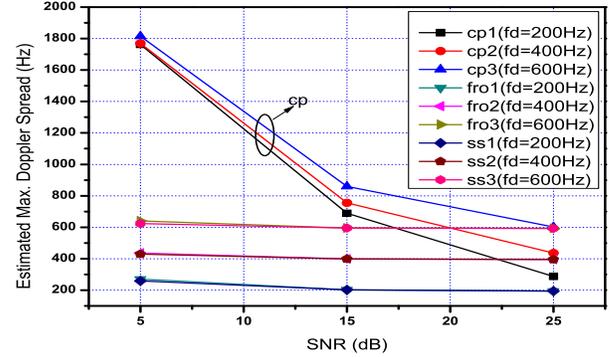}
\caption{Performance comparison for the CP-based \cite{Cai03},
Frobenius-norm-based (\ref{eqn:etadefnew}) and subspace-based
(\ref{eqn:etadefnewrefine}) algorithms when a 20ms frame is used and
$\beta=1$.} \label{fig:fig1}
\end{figure}

\section{Simulation Results}
\label{sec:simulation} The performance of the proposed algorithm is
evaluated for an OFDM system with $BW=5$ MHz ($T=1/BW=200$ ns),
$N=512$, $L_{cp}=64$ and $P=64$. ITU Vehicular A Channels
\cite{ITUR1225} is adopted, which consists of six individually faded
taps with relative delays as $[ 0, 310, 710, 1090, 1730, 2510 ]$ ns
and average power as $[ 0, -1, -9, -10, -15, -20 ]$ dB. The classic
Doppler spectrum, i.e., Jakes' spectrum \cite{Stee92}, is applied to
generate the Rayleigh fading channel.

In Fig.\ref{fig:fig1}, a CP-based algorithm reported in \cite{Cai03}
and (\ref{eqn:etadefnew}), which is based on the Frobenius norm, are
compared with the proposed subspace-based algorithm
(\ref{eqn:etadefnewrefine}) with $\beta=1$. A 20ms frame including
194 OFDM symbols is used to generate the statistics. $f_d=200$,
$400$ and $600$ Hz are tested under a range of SNR's, respectively.
Apparently, the CP-based algorithm fails for all $f_d$'s when the
SNR is below 20 dB, meanwhile (\ref{eqn:etadefnew}) and
(\ref{eqn:etadefnewrefine}) work very well for all $f_d$'s and SNR's
but with a moderate positive bias for $SNR=5$ dB. In fact, when
$SNR=5$ dB, resorting to (\ref{eqn:rholowbound}) and
(\ref{eqn:rhorandrho}), the lower bound of the bias terms $\rho$ and
$\rho_r$ are 0.0395 for (\ref{eqn:etadefnew}) and 0.0099 for
(\ref{eqn:etadefnewrefine}), respectively. And when $f_d=600$ Hz,
according to (\ref{eqnarray:xi0def}) and (\ref{eqnarray:xibetadef}),
$\xi(0)=0.9938$ and $\xi(\beta)=0.9476$. Thus, the relative errors
of $\eta$ are 0.0007 and 0.0000 for (\ref{eqn:etadefnew}) and
(\ref{eqn:etadefnewrefine}), respectively, which almost have no
effect on the estimation of the maximum Doppler spread. Therefore,
(\ref{eqn:etadefnew}) and (\ref{eqn:etadefnewrefine}) show almost
the same performance when SNR's are above 5 dB.

\begin{figure}[!t]
\centering
\includegraphics[width=3.7in, height=2.3in]{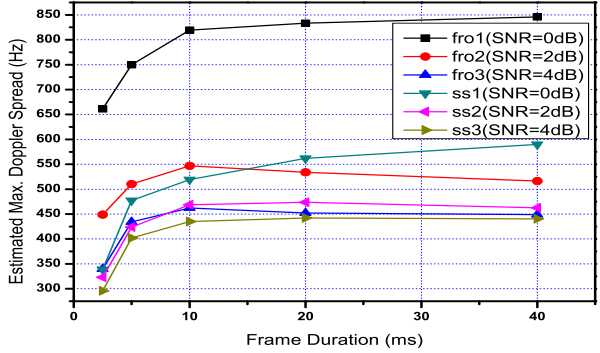}
\caption{Performance comparison between the Frobenius-norm-based
(\ref{eqn:etadefnew}) and subspace-based (\ref{eqn:etadefnewrefine})
algorithms when $f_d=400$ Hz and $\beta=1$.} \label{fig:fig2}
\end{figure}

Fig.\ref{fig:fig2} shows the performance comparison between the
Frobenus-norm-based (\ref{eqn:etadefnew}) and subspace-based
(\ref{eqn:etadefnewrefine}) in the low SNR regime, specifically,
below 5 dB, to emphasize the capability of noise depression of the
latter. Different frame durations are used to obtain TACF. From the
figure we can find (\ref{eqn:etadefnewrefine}) outperforms
(\ref{eqn:etadefnew}) for all the SNR's and frame durations,
although both of them over-estimate the maximum Doppler spread due
to the non-negligible noise bias term in the low SNR regime, which
is also the reason why increasing the length of observation record
does not help to decrease the bias in this regime.

\begin{figure}[!t]
\centering
\includegraphics[width=3.7in, height=2.3in]{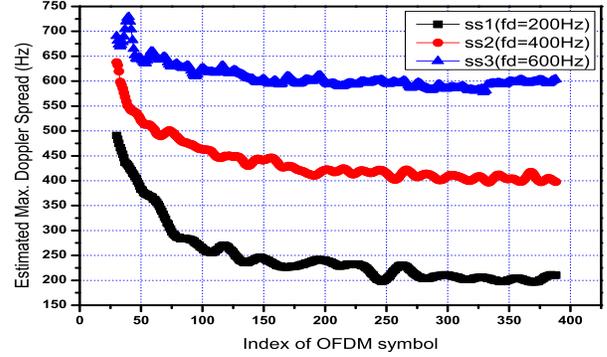}
\caption{The convergence of the proposed subspace-tracking-based
algorithm when $SNR=15$ dB and $\beta=1$.} \label{fig:fig3}
\end{figure}

The convergence of the proposed subspace-tracking-based algorithm is
shown in Fig.\ref{fig:fig3}. Three different maximum Doppler spreads
are tested, i.e., $f_d=200$, $400$ and $600$ Hz, when $SNR=15$ dB.
The curves are plotted from the thirtieth OFDM symbol. It is
observed from the plot that all the three curves converge to their
true values after numbers of OFDM symbols and, further, the higher
the Doppler spread is, the faster the curve converges. This is due
to the subspace is updating faster when the Doppler spread is
higher. In addition, after converging, the estimated maximum Doppler
spread is fluctuating around its true value in a small range, hence
additional time-averaging can be employed to smooth the curve.

\section{Conclusions}
\label{sec:conclusion} In this paper, we propose a
subspace-tracking-based maximum Doppler spread estimation algorithm
which is applicable to OFDM systems with the comb-type pilot
pattern. It enjoys three main advantages: i) alleviating the noise
term; ii) reducing the computation complexity; iii) tracking the
drifting delay subspace. Through simulations, the performance of the
proposed algorithm is demonstrated to outperform the CP-based
algorithm \cite{Cai03}. Moreover, since the proposed algorithm is
based on the subspace tracking, it can be easily integrated into the
channel estimator equipped with the subspace tracker
\cite{Sime04}\cite{Ragh07}, which lends a broad application promise
to it.

\setlength{\arraycolsep}{5pt}

\bibliographystyle{IEEEtran}
\bibliography{IEEEabrv,myBibs}

\end{document}